\PassOptionsToPackage{unicode}{hyperref}
\PassOptionsToPackage{hyphens}{url}
\documentclass[
  11pt,
]{article}
\usepackage{xcolor}
\usepackage[margin=2.5cm]{geometry}
\usepackage{amsmath,amssymb}
\usepackage{iftex}
\ifPDFTeX
  \usepackage[T1]{fontenc}
  \usepackage[utf8]{inputenc}
  \usepackage{textcomp} 
\else 
  \usepackage{unicode-math} 
  \defaultfontfeatures{Scale=MatchLowercase}
  \defaultfontfeatures[\rmfamily]{Ligatures=TeX,Scale=1}
\fi
\usepackage{lmodern}
\ifPDFTeX\else
\fi
\IfFileExists{upquote.sty}{\usepackage{upquote}}{}
\IfFileExists{microtype.sty}{
  \usepackage[]{microtype}
  \UseMicrotypeSet[protrusion]{basicmath} 
}{}
\makeatletter
\@ifundefined{KOMAClassName}{
  \IfFileExists{parskip.sty}{%
    \usepackage{parskip}
  }{
    \setlength{\parindent}{0pt}
    \setlength{\parskip}{6pt plus 2pt minus 1pt}}
}{
  \KOMAoptions{parskip=half}}
\makeatother
\usepackage{graphicx}
\makeatletter
\newsavebox\pandoc@box
\newcommand*\pandocbounded[1]{
  \sbox\pandoc@box{#1}%
  \Gscale@div\@tempa{\textheight}{\dimexpr\ht\pandoc@box+\dp\pandoc@box\relax}%
  \Gscale@div\@tempb{\linewidth}{\wd\pandoc@box}%
  \ifdim\@tempb\p@<\@tempa\p@\let\@tempa\@tempb\fi
  \ifdim\@tempa\p@<\p@\scalebox{\@tempa}{\usebox\pandoc@box}%
  \else\usebox{\pandoc@box}%
  \fi%
}
\def\fps@figure{htbp}
\makeatother
\NewDocumentCommand\citeproctext{}{}
\NewDocumentCommand\citeproc{mm}{%
  \begingroup\def\citeproctext{#2}\cite{#1}\endgroup}
\makeatletter
 \let\@cite@ofmt\@firstofone
 \def\@biblabel#1{}
 \def\@cite#1#2{{#1\if@tempswa , #2\fi}}
\makeatother
\newlength{\cslhangindent}
\newlength{\csllabelwidth}
\newenvironment{CSLReferences}[2] 
 {\begin{list}{}{%
  \setlength{\itemindent}{0pt}
  \setlength{\leftmargin}{0pt}
  \setlength{\parsep}{0pt}
  \ifodd #1
   \setlength{\leftmargin}{\cslhangindent}
   \setlength{\itemindent}{-1\cslhangindent}
  \fi
  \setlength{\itemsep}{#2\baselineskip}}}
 {\end{list}}
\usepackage{calc}

\newcommand{\CSLLeftMargin}[1]{\parbox[t]{\csllabelwidth}{\strut#1\strut}}
\newcommand{\CSLRightInline}[1]{\parbox[t]{\linewidth - \csllabelwidth}{\strut#1\strut}}

\providecommand{\tightlist}{%
  \setlength{\itemsep}{0pt}\setlength{\parskip}{0pt}}
\usepackage{booktabs}
\usepackage{longtable}
\usepackage{array}
\usepackage{multirow}
\usepackage{wrapfig}
\usepackage{float}
\usepackage{colortbl}
\usepackage{pdflscape}
\usepackage{tabu}
\usepackage{threeparttable}
\usepackage{threeparttablex}
\usepackage{makecell}
\usepackage{xcolor}
\usepackage{tabularx}
\usepackage{xltabular}
\usepackage[normalem]{ulem}
\usepackage{bookmark}
\IfFileExists{xurl.sty}{\usepackage{xurl}}{} 
\makeatletter
\@ifundefined{xmpquote}{}{}
\makeatother
\hypersetup{
  pdftitle={Global maps of travel time to emergency and tertiary hospitals},
  pdfauthor={Johan Emilsson},
  hidelinks,
  pdfcreator={LaTeX via pandoc}}

\title{Global maps of travel time to emergency and tertiary hospitals}
\author{Johan Emilsson\footnote{Independent researcher. Correspondence:
  \href{mailto:johan.emilsson@eupi.org}{\nolinkurl{johan.emilsson@eupi.org}}}}
\date{10 September 2026}

\begin{document}
\maketitle
\begin{abstract}
Global estimates of access to health care treat every facility alike.
Drive time was routed on the OpenStreetMap network from all 217,841 of
its hospitals and 15-, 30- and 60-minute isochrones dissolved for nested
capability tiers, weighted with GHS-POP 2025. 96.2\% of the
world\textquotesingle s population lives within an
hour\textquotesingle s drive of a hospital, 78.5\% of a hospital with a
documented emergency department, and 40.3\% of one with a helipad or
university name marking tertiary care. The 1.45 billion people within an
hour of a hospital but not of an emergency hospital live mainly in South
and East Asia and Nigeria; the 312 million beyond an hour of any
hospital lie in a belt from the Sahel to Madagascar. Defining the
tertiary tier by helipad alone leaves its estimate unchanged, and the
isochrone construction shifts the 60-minute figures by a few percent.
Counting every facility as equivalent overstates access to time-critical
care.
\end{abstract}

\section{Introduction}\label{introduction}

Physical access to a hospital is a precondition for treating the
conditions that kill quickly: severe injury, obstetric haemorrhage,
myocardial infarction, stroke and sepsis. Outcomes for each worsen with
time to definitive care, even if the shape of the time-to-outcome curve
is
debated\textsuperscript{\citeproc{ref-harmsen2015}{1}--\citeproc{ref-clark2025}{5}}.
Health-system benchmarks are therefore expressed in travel time: 80\% of
a population within two hours of a facility able to perform the
bellwether procedures\textsuperscript{\citeproc{ref-meara2015}{6}}, a
60-minute "golden hour" to a high-level trauma
centre\textsuperscript{\citeproc{ref-branas2005}{7},\citeproc{ref-medrano2023}{8}},
time to a thrombectomy-capable stroke
centre\textsuperscript{\citeproc{ref-sarraj2020}{9},\citeproc{ref-asif2024}{10}},
and 120 minutes to primary percutaneous coronary
intervention\textsuperscript{\citeproc{ref-ibanez2018}{11}}.

The reference global estimate is that of Weiss and
colleagues\textsuperscript{\citeproc{ref-weiss2020}{12}}, who combined
facility locations from Google Maps, OpenStreetMap (OSM) and published
inventories with a one-kilometre friction
surface\textsuperscript{\citeproc{ref-weiss2018}{13}} and found that
91.1\% of the world\textquotesingle s population could reach a hospital
or clinic within one hour by motorised transport. One limitation named
by its authors and one inherent in the method motivate this study.
First, it treats every hospital and clinic as equivalent, so that
"travel times to more specialized services could be much greater than
our maps indicate"\textsuperscript{\citeproc{ref-weiss2020}{12}}.
Second, a friction surface lets travel proceed across any land cell at
some speed, which makes it less sensitive than a routed estimate to the
completeness and connectivity of the mapped road network. Regional
comparisons find that raster and network methods can classify
substantially different populations as
underserved\textsuperscript{\citeproc{ref-delamater2012}{14}--\citeproc{ref-macharia2024}{16}},
and field replication in African cities shows that both tend to be
optimistic\textsuperscript{\citeproc{ref-banke2021}{17},\citeproc{ref-whitaker2022}{18}}.

Capability-differentiated access has been studied at national and
continental scale: emergency hospitals in sub-Saharan Africa, where 29\%
of the population lived more than two hours from
one\textsuperscript{\citeproc{ref-ouma2018}{19}}; major hospitals
assumed capable of essential
surgery\textsuperscript{\citeproc{ref-juran2018}{20}}; hospitals versus
health centres in public-facility
databases\textsuperscript{\citeproc{ref-maina2019}{21},\citeproc{ref-falchetta2020}{22}};
trauma
centres\textsuperscript{\citeproc{ref-branas2005}{7},\citeproc{ref-medrano2023}{8},\citeproc{ref-smedley2019}{23}},
thrombectomy centres\textsuperscript{\citeproc{ref-sarraj2020}{9}} and
24-hour emergency
departments\textsuperscript{\citeproc{ref-mcgaughey2024}{24}} in
high-income countries. Wu and colleagues recently mapped global access
to essential services, health facilities among them, with a 30-metre
friction surface, again without distinguishing facility
levels\textsuperscript{\citeproc{ref-wu2025}{25}}. No global estimate
known to the author restricts attention to hospitals, distinguishes them
by documented emergency capability, and routes on the road network.

This paper provides one. Drive-time isochrones were computed on the OSM
road network for every hospital in OSM and dissolved for three nested
tiers: any hospital, hospitals tagged as having an emergency department,
and emergency hospitals with a marker of tertiary capability (a helipad
or runway on the grounds, or a university or teaching hospital name).
Coverage at 15, 30 and 60 minutes is overlaid with the 2025 Global Human
Settlement population
grid\textsuperscript{\citeproc{ref-schiavina2023}{26}} and reported by
country, region and income group. The any-hospital tier is benchmarked
against the published friction-surface
raster\textsuperscript{\citeproc{ref-mapraster}{27}} re-weighted to the
same population grid, and the share of hospitals with a documented
emergency department is treated as a completeness indicator that
conditions the apparent gap between tiers.

\section{Data and methods}\label{data-and-methods}

\subsection{Hospital inventory and capability
tiers}\label{hospital-inventory-and-capability-tiers}

Hospitals are every node or closed way tagged \texttt{amenity=hospital}
in the OpenStreetMap planet file of 10 August
2026\textsuperscript{\citeproc{ref-osm}{28}}; polygons were reduced to
centroids. The extraction yielded 217,841 hospital records (216,889
distinct OSM ids; a few objects appear as both node and polygon), 91.5\%
of them named. Clinics, doctors\textquotesingle{} practices and
pharmacies, which Weiss and
colleagues\textsuperscript{\citeproc{ref-weiss2020}{12}} and
healthsites.io\textsuperscript{\citeproc{ref-saameli2018}{29}} group
with hospitals, were excluded: the question here is access to the level
of care a hospital represents.

Three nested tiers were defined from OSM attributes (Table
\ref{tab:tiers}):

\begin{itemize}
\tightlist
\item
  \textbf{Any hospital}: all \texttt{amenity=hospital} features
  (217,841).
\item
  \textbf{Emergency}: hospitals also tagged \texttt{emergency=yes}
  (25,422, 11.7\%). The OSM wiki defines the tag as indicating an
  emergency department. Its absence means only that no contributor
  recorded one.
\item
  \textbf{Emergency with tertiary marker}: emergency hospitals that
  either contain a helipad, heliport or runway (\texttt{aeroway=*})
  within the hospital polygon, or whose name, English name, operator,
  official or alternative name contains a root for "university" or
  "teaching" from a list covering about fifty languages and scripts,
  including the French abbreviations CHU, CHUV and CHRU (6,715, 3.1\%).
  Address matches such as "University Avenue" were excluded. The marker
  is a proxy for specialist services (neurosurgery, interventional
  radiology, cardiac surgery) that most health systems concentrate in
  teaching hospitals and in hospitals receiving helicopter transfers.
  Helipad presence is close to complete within the emergency-tagged set:
  in a MapRoulette challenge created by the author in 2016 and
  maintained until 2026, whose task set is every OSM hospital tagged
  \texttt{emergency=yes}, all 22,502 tasks were reviewed against aerial
  imagery; 3,497 hospitals already had a mapped helipad, 1,741 had one
  added, and 17,264 were confirmed to have
  none\textsuperscript{\citeproc{ref-maproulette}{30}}. The challenge
  accepted helipads on or near the hospital, whereas the tier requires
  the helipad inside the hospital polygon (or within 100 m of a hospital
  node), so a helipad mapped just outside the grounds is not credited.
  The university-name marker was not reviewed in the same way.
\end{itemize}

A fourth tier, \textbf{helipad}, is defined by a helipad on the grounds
alone, with no requirement for the emergency tag (8,287 hospitals). It
is not nested in the others and was built after the main run by
dissolving the per-hospital polygons of the qualifying hospitals; it
serves to test whether the tertiary estimate depends on emergency
tagging.

\begin{table}[!h]
\centering
\caption{\label{tab:tiers}Capability tiers used in this study. The first three are nested; the helipad tier overlaps the emergency tiers but does not require the emergency tag.}
\centering
\fontsize{9}{11}\selectfont
\begin{tabular}[t]{l>{\raggedright\arraybackslash}p{7.5cm}ll}
\toprule
Tier & OSM definition & Hospitals & Share\\
\midrule
Any hospital & amenity=hospital (node or closed way) & 217,841 & 100.0\%\\
Emergency (tagged) & amenity=hospital and emergency=yes & 25,422 & 11.7\%\\
Emergency + tertiary marker & emergency=yes and (aeroway helipad/heliport/runway within hospital grounds, or university/teaching-hospital name root) & 6,715 & 3.1\%\\
Helipad (not nested) & aeroway helipad/heliport/runway within hospital grounds, any emergency status & 8,287 & 3.8\%\\
\bottomrule
\end{tabular}
\end{table}

\subsection{Road network and drive-time
isochrones}\label{road-network-and-drive-time-isochrones}

The road network from the same planet file was converted to a routable
graph with osm2po 5.5.11\textsuperscript{\citeproc{ref-osm2po}{31}}
using its car profile: motorway, trunk, primary, secondary, tertiary,
residential, unclassified and "road" classes and their link roads, at
default free-flow speeds of 120, 90, 70, 60, 40, 40 and 50 km/h,
overridden by \texttt{maxspeed} where tagged. Service roads, tracks,
living streets, paths, footways and ways closed to motor vehicles were
excluded; one-way restrictions were respected; ferries were not
routable. Edge costs are hours of free-flow driving. No allowance was
made for congestion, road condition, borders or time of day, so travel
times are optimistic in the same sense as the "optimal" scenario of
Weiss and colleagues\textsuperscript{\citeproc{ref-weiss2020}{12}}.

Each hospital was snapped to its nearest graph vertex; hospitals sharing
a vertex share an isochrone. Vertices were grouped into one-degree cells
and, per cell, the subgraph within a two-degree buffer (about 220 km at
the equator, more than 60 minutes at 120 km/h) was loaded into memory.
Travel times from all origins in a cell were computed by bounded
multi-source Dijkstra
search\textsuperscript{\citeproc{ref-dijkstra1959}{32}} in
SciPy\textsuperscript{\citeproc{ref-virtanen2020}{33}}, truncated at 60
minutes. The search runs outbound from the hospital; treating travel
time as symmetric is an approximation that one-way streets violate by
minutes near the hospital.

\begin{figure}[H]

{\centering \includegraphics[width=1\linewidth]{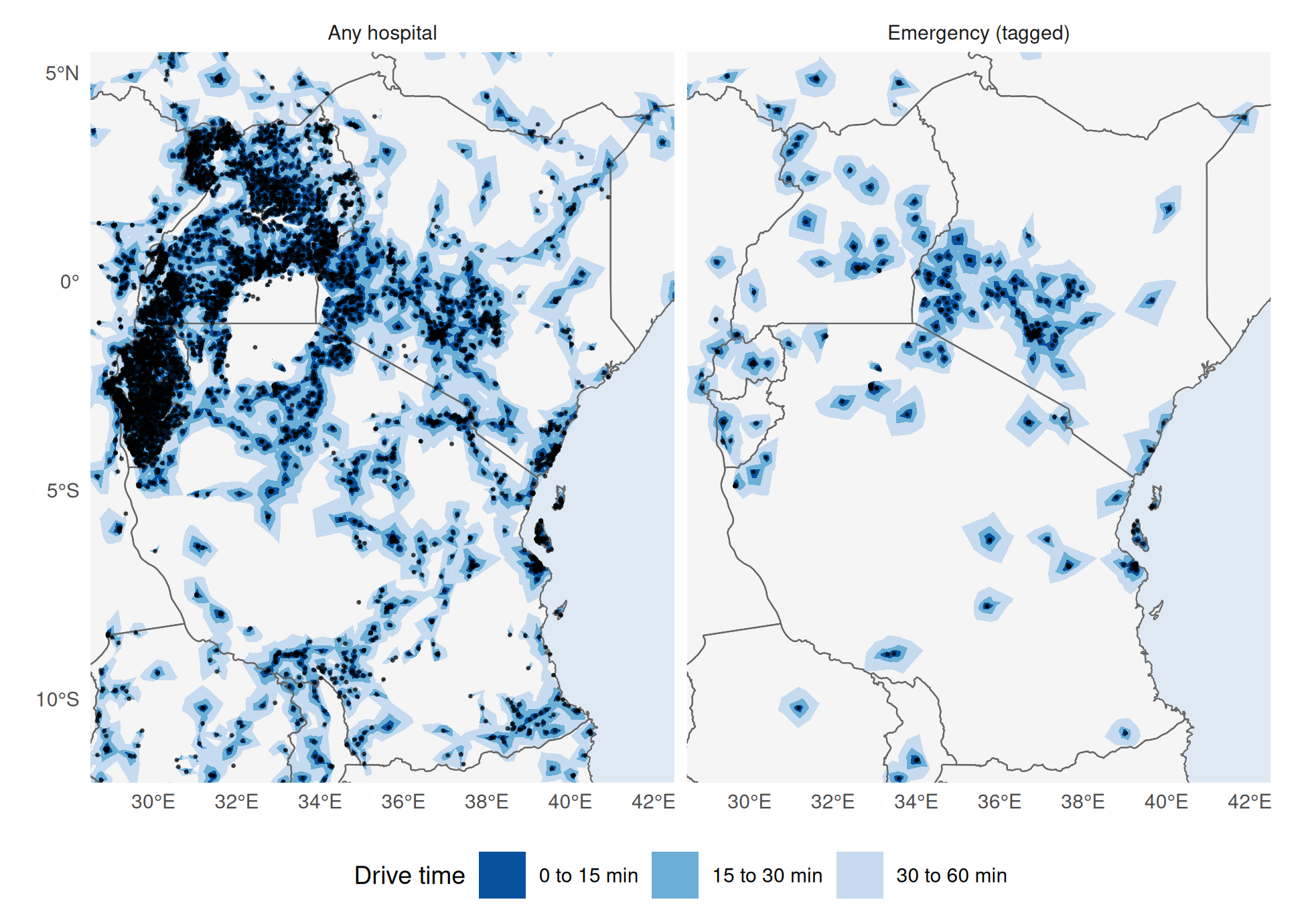} 

}

\caption{Method illustration for East Africa. Dissolved drive-time bands (0 to 15, 15 to 30 and 30 to 60 minutes) around any OpenStreetMap hospital (left) and around hospitals tagged with an emergency department (right). Points mark hospital locations. Bands are computed as concave hulls of the road-network vertices reachable within the time budget and dissolved across hospitals.}\label{fig:example-fig}
\end{figure}

For each origin and band (15, 30, 60 minutes) the reached vertices were
converted to a polygon with the GEOS concave hull at concavity ratio
0.85\textsuperscript{\citeproc{ref-park2012}{34}}, where at least three
vertices were reached; on this scale 1 is the convex hull and 0 the most
concave shape, so the polygons are close to convex. Hulls were chosen
over buffered edges for tractability at planet scale (649,821 polygons);
their consequences are discussed under limitations. 216,545, 216,629 and
216,647 hospitals received a 15-, 30- and 60-minute polygon; the
remaining 1,194 (0.5\%) snapped to vertices disconnected from any
drivable road, typically on small islands or where no roads of the
included classes are mapped. Isochrones were dissolved by tier and band
into nine global coverage polygons (Figure \ref{fig:example-fig}),
simplified to about 50 m. Cumulative coverage (everything within the
band) is used throughout.

Two plausibility checks were run on every per-hospital polygon. No
polygon exceeds the area of the disc reachable at the maximum speed (30,
60 and 120 km radius for the three bands); the median 60-minute polygon
covers 9,606 km², equivalent to a disc of radius 55 km, and the largest
28,142 km². The distance from a hospital to the centroid of its polygon,
which exposes snapping to a distant vertex, exceeds 50 km for 10 of the
216,545 15-minute polygons. Between 0.9\% (15 minutes) and 2.7\% (60
minutes, tertiary tier) of the dissolved polygons\textquotesingle{} area
lies over water or outside every country polygon.

To quantify the effect of the hull construction, the routing stage was
rerun once with three concavity ratios (0.85, 0.5 and 0.3) for every
source, and, for a quasi-random 2\% sample of sources (every fiftieth
graph vertex id; 4,319 hospitals, of which 516 emergency-tagged and 128
with a tertiary marker; counts in Table \ref{tab:sens}), with a non-hull
reference polygon: the union of 1.5 km buffers around all reached
vertices, computed on a 100 m grid in a local equidistant projection. In
this rerun hulls were computed on at most 50,000 randomly subsampled
reached vertices per source to bound memory; the 0.85 hulls reproduce
the main run\textquotesingle s coverage figures to one decimal place.
The buffer polygon is a conservative reference rather than ground truth:
it excludes anyone more than 1.5 km from a vertex of the drivable graph,
including people served by service roads, tracks or footpaths that the
graph omits.

\subsection{Population, land area and country
attribution}\label{population-land-area-and-country-attribution}

Population is GHS-POP release R2023A, epoch 2025, at 30 arc-seconds in
WGS84\textsuperscript{\citeproc{ref-schiavina2023}{26},\citeproc{ref-freire2016}{35}},
which disaggregates census estimates by satellite-detected built-up
area; the grid totals 8.19 billion people. Country boundaries and UN
region attributes are from Natural Earth admin-0, 1:10 million, version
5.1.1\textsuperscript{\citeproc{ref-naturalearth}{36}}. Income groups
are the World Bank classification for fiscal year 2027, in effect from 1
July 2026, retrieved from the World Bank API on 5 September 2026 and
joined by ISO 3166-1 alpha-3
code\textsuperscript{\citeproc{ref-worldbank2026}{37}}. Territories the
World Bank does not classify were assigned as follows: dependencies take
the group of their sovereign state; Taiwan is assigned to high income,
its GNI per capita being more than twice the high-income threshold;
Somaliland takes Somalia\textquotesingle s group, Northern Cyprus and
the UN buffer zone Cyprus\textquotesingle s, and Western Sahara
Morocco\textquotesingle s. Only uninhabited territories remain
unclassified. Countries were rasterised to the population grid by cell
centre, with an all-touched pass filling coastal and small-island cells;
coverage polygons were rasterised by cell centre. Population and
geodesic land area within each tier and band were then summed by
country. 34.6 million people (0.42\%) on cells outside every Natural
Earth polygon enter global totals but not country tables. Hospitals were
attributed to countries by point-in-polygon with a nearest-polygon
fallback within 0.5 degrees.

\subsection{Benchmark against friction-surface travel
time}\label{benchmark-against-friction-surface-travel-time}

The Malaria Atlas Project distributes the 2020 raster of motorised
travel time to the nearest hospital or clinic underlying Weiss and
colleagues\textsuperscript{\citeproc{ref-weiss2020}{12},\citeproc{ref-mapraster}{27}},
at 30 arc-seconds between 60 degrees south and 85 degrees north. It was
resampled to the GHS-POP grid by nearest neighbour, and the population
within 10, 30, 60 and 120 minutes was summed per country with the 2025
grid. Holding population constant isolates the difference between access
models (facility definition, network versus friction surface, data
vintage 2019 versus 2026) from the difference between population grids.
Population was also cross-tabulated cell by cell between
friction-surface class (60 minutes or less, more) and membership of the
any-hospital and emergency 60-minute polygons.

\subsection{Indicators}\label{indicators}

Per country the analysis reports the share of population within each
band of each tier; the \emph{capability gap}, the population within 60
minutes of any hospital but not of an emergency-tagged hospital; the
\emph{documented-emergency share}, the fraction of hospitals tagged
\texttt{emergency=yes}; and hospitals per 100,000 population. Country
statistics cover the 162 countries and territories with at least one
million inhabitants; the benchmark covers those with friction-surface
data for at least 90\% of their population. Group aggregates are
population-weighted. Processing was done in Python (GDAL, rasterio,
GeoPandas) and R.

\section{Results}\label{results}

\subsection{The hospital inventory}\label{the-hospital-inventory}

Mapped hospital density varies 3.4-fold between income groups (Table
\ref{tab:income}), from 4.4 per 100,000 people in high-income to 1.3 in
low-income countries. India alone contributes 55,634 hospitals, a
quarter of the total. The documented-emergency share is 11.7\% globally,
19.8\% in high-income and 3.7\% in low-income countries, and ranges
between countries with at least 20 mapped hospitals from 0.0\% (North
Korea) to 72\% (Costa Rica); India, Japan and Nigeria are all below
3.1\%. Variation of that size is not a credible description of
hospitals; it describes tagging practice (Section \ref{sec:gap}).

\begin{table}[!h]
\centering
\caption{\label{tab:income}Hospital inventory and 60-minute drive-time coverage by World Bank income group (fiscal year 2027 classification), countries with at least one million inhabitants. The last column gives the share within 60 minutes of any hospital or clinic according to the friction-surface raster of Weiss et al., re-weighted to the 2025 population grid.}
\centering
\fontsize{8}{10}\selectfont
\begin{tabular}[t]{lrrrrrrrrrr}
\toprule
\multicolumn{5}{c}{ } & \multicolumn{2}{c}{Share of hospitals} & \multicolumn{4}{c}{Population within 60 min drive} \\
\cmidrule(l{3pt}r{3pt}){6-7} \cmidrule(l{3pt}r{3pt}){8-11}
Income group & N & \makecell[r]{Pop.\\(bn)} & Hospitals & \makecell[r]{per\\100k} & \makecell[r]{Emerg.\\share} & \makecell[r]{Tert.\\share} & \makecell[r]{Any\\hospital} & \makecell[r]{Emerg.\\hospital} & \makecell[r]{Tert.\\marker} & \makecell[r]{Friction\\surface}\\
\midrule
High income & 50 & 1.40 & 61,636 & 4.4 & 19.8\% & 9.2\% & 99.7\% & 96.8\% & 86.7\% & 99.7\%\\
Upper middle income & 44 & 3.07 & 62,397 & 2.0 & 14.6\% & 1.4\% & 98.1\% & 83.7\% & 48.1\% & 97.1\%\\
Lower middle income & 42 & 2.92 & 83,077 & 2.8 & 4.3\% & 0.1\% & 97.8\% & 75.4\% & 18.8\% & 97.3\%\\
Low income & 26 & 0.75 & 9,771 & 1.3 & 3.7\% & 0.2\% & 76.6\% & 35.8\% & 5.6\% & 88.1\%\\
\bottomrule
\end{tabular}
\end{table}

\subsection{Global coverage by tier and
band}\label{global-coverage-by-tier-and-band}

\begin{table}[!h]
\centering
\caption{\label{tab:global}Global population and land area within drive-time bands of hospitals, by capability tier. Population from GHS-POP 2025 (8.19 billion). Land area is the area of grid cells attributed to a country; for the helipad tier it is the equal-area extent of the dissolved polygon.}
\centering
\resizebox{\ifdim\width>\linewidth\linewidth\else\width\fi}{!}{
\fontsize{9}{11}\selectfont
\begin{tabular}[t]{llrrrrr}
\toprule
Tier & Band & Population (bn) & Share & Not covered (bn) & Land area (M km2) & Land share\\
\midrule
 & 15 min & 6.16 & 75.2\% & 2.03 & 14.6 & 9.9\%\\

 & 30 min & 7.32 & 89.4\% & 0.87 & 34.2 & 23.3\%\\

\multirow[t]{-2}{*}[\normalbaselineskip]{\raggedright\arraybackslash Any hospital} & 60 min & 7.88 & 96.2\% & 0.31 & 58.0 & 39.4\%\\
\cmidrule{1-7}
 & 15 min & 3.44 & 42.0\% & 4.75 & 5.7 & 3.9\%\\

 & 30 min & 4.75 & 57.9\% & 3.45 & 17.3 & 11.7\%\\

\multirow[t]{-2}{*}[\normalbaselineskip]{\raggedright\arraybackslash Emergency (tagged)} & 60 min & 6.43 & 78.5\% & 1.76 & 36.4 & 24.8\%\\
\cmidrule{1-7}
 & 15 min & 1.62 & 19.8\% & 6.57 & 2.6 & 1.7\%\\

 & 30 min & 2.36 & 28.8\% & 5.84 & 8.1 & 5.5\%\\

\multirow[t]{-2}{*}[\normalbaselineskip]{\raggedright\arraybackslash Emergency + tertiary marker} & 60 min & 3.30 & 40.3\% & 4.89 & 16.8 & 11.4\%\\
\cmidrule{1-7}
 & 15 min & 1.76 & 21.4\% & 6.44 & 3.0 & \\

 & 30 min & 2.47 & 30.2\% & 5.72 & 9.3 & \\

\multirow[t]{-2}{*}[\normalbaselineskip]{\raggedright\arraybackslash Helipad (not nested)} & 60 min & 3.30 & 40.3\% & 4.89 & 19.5 & \\
\bottomrule
\end{tabular}}
\end{table}

96.2\% of the world\textquotesingle s population (7.88 billion) lives
within a 60-minute free-flow drive of a mapped hospital, 89.4\% within
30 minutes and 75.2\% within 15 (Table \ref{tab:global}). For hospitals
with a documented emergency department the shares are 78.5\%, 57.9\% and
42.0\%; for emergency hospitals with a tertiary marker 40.3\%, 28.8\%
and 19.8\%. In absolute terms, 312 million people live more than an hour
from any mapped hospital, 1.76 billion from any documented emergency
hospital and 4.89 billion from any hospital with a tertiary marker. Land
coverage is far lower than population coverage: 39.4\% of the land
surface lies within 60 minutes of any hospital and 11.4\% within 60
minutes of a tertiary-marker hospital.

The tertiary marker is dominated by helipads: of the 6,650 distinct
emergency hospitals with a marker (6,715 records), 5,685 qualify by
helipad alone, 623 by name alone and 342 by both. Both markers are
frequently present on hospitals that lack the emergency tag: 2,260 of
the 8,287 hospitals with a helipad (27\%) and 1,652 of the 2,617 with a
university name (63\%) are not tagged \texttt{emergency=yes}. The
helipad tier, which ignores the emergency tag, nevertheless covers
40.3\% of the population within 60 minutes, the same as the
emergency-plus-marker tier, and slightly more at 30 and 15 minutes
(30.2\% and 21.4\%). Dropping the university-named hospitals and adding
the untagged helipad hospitals therefore leaves the tertiary estimate
unchanged, so this tier, unlike the emergency tier, is insensitive to
the emergency-tag requirement.

\subsection{Geography of the gaps}\label{geography-of-the-gaps}

\begin{figure}[p]

{\centering \includegraphics[width=1\linewidth]{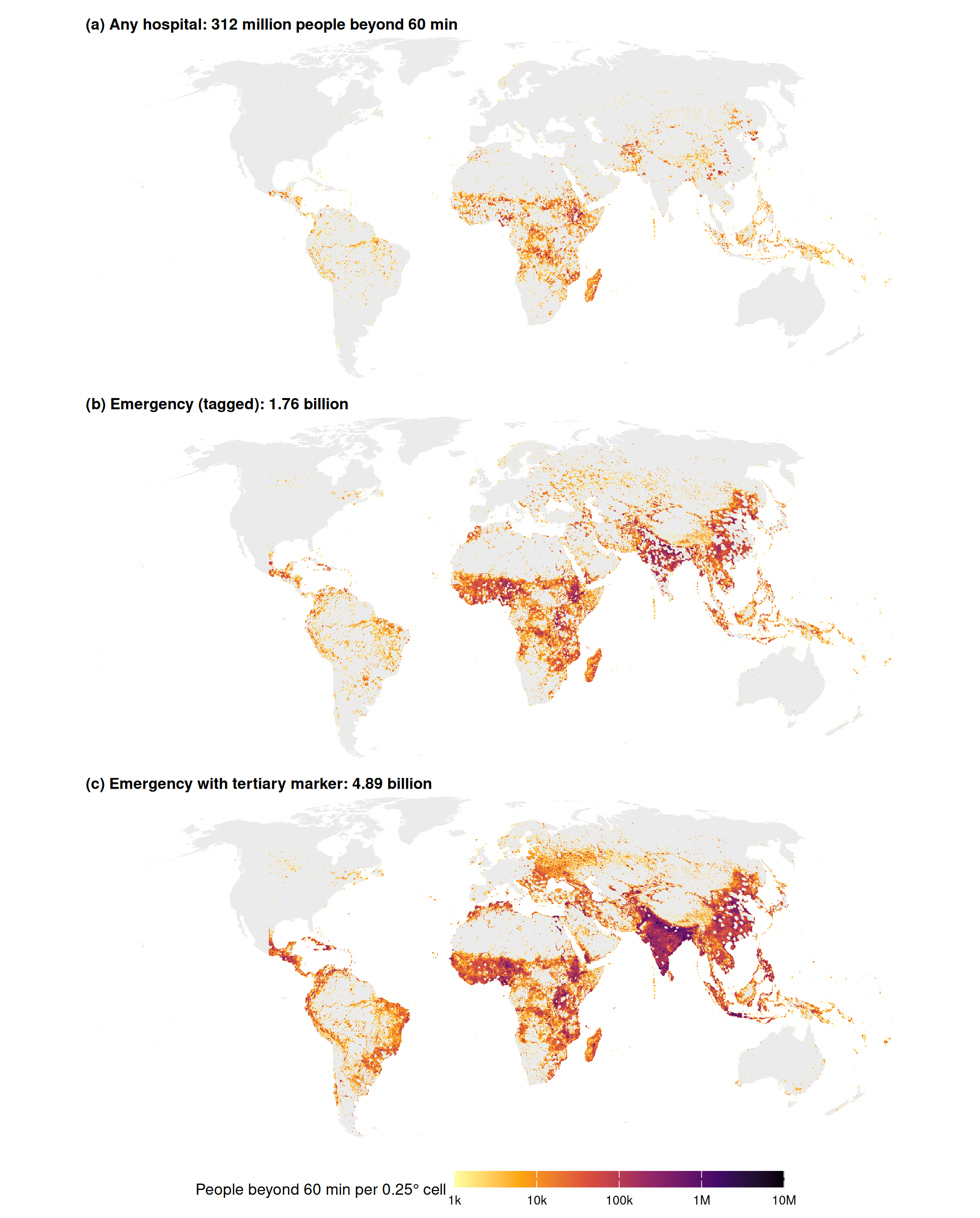} 

}

\caption{Population living more than 60 minutes' drive from (a) any hospital, (b) any hospital tagged with an emergency department, and (c) any emergency hospital with a tertiary marker, summed on a 0.25 degree grid (GHS-POP 2025). Cells with fewer than 1,000 uncovered residents are not shown. Robinson projection.}\label{fig:maps}
\end{figure}

The population beyond 60 minutes of any hospital forms a belt from the
Sahel through Central and East Africa to Madagascar, with secondary
concentrations in Afghanistan, Papua New Guinea, the Democratic
People\textquotesingle s Republic of Korea and the interior of Brazil
and Peru (Figure \ref{fig:maps}a). Africa holds 18.6\% of the
world\textquotesingle s population and 66.1\% of the people beyond an
hour of any hospital; 86.4\% of Africans are within an hour of a mapped
hospital, 54.0\% of an emergency-tagged one and 23.2\% of one with a
tertiary marker. The Democratic Republic of the Congo and Ethiopia alone
account for 82 million people beyond an hour (Table \ref{tab:worst});
both have 0.6 and 0.3 mapped hospitals per 100,000 inhabitants.

\begin{table}[!h]
\centering
\caption{\label{tab:worst}The twenty countries with the largest population more than 60 minutes' drive from any mapped hospital.}
\centering
\fontsize{8}{10}\selectfont
\begin{tabular}[t]{lrrrrrrrr}
\toprule
Country & \makecell[r]{Pop.\\(M)} & Hospitals & \makecell[r]{per\\100k} & \makecell[r]{Beyond 60 min\\of any (M)} & \makecell[r]{Share\\beyond} & \makecell[r]{Within 60,\\emergency} & \makecell[r]{Within 60,\\tertiary} & \makecell[r]{Friction\\surface, 60}\\
\midrule
Dem. Rep. Congo & 108.8 & 623 & 0.6 & 42.0 & 38.6\% & 25.8\% & 0.0\% & 96.8\%\\
Ethiopia & 132.2 & 405 & 0.3 & 40.0 & 30.2\% & 23.3\% & 4.4\% & 89.8\%\\
China & 1,420.3 & 12,922 & 0.9 & 23.2 & 1.6\% & 78.8\% & 50.8\% & 96.2\%\\
Sudan & 50.8 & 375 & 0.7 & 19.4 & 38.1\% & 26.1\% & 0.0\% & 62.3\%\\
Nigeria & 233.5 & 4,254 & 1.8 & 17.3 & 7.4\% & 56.2\% & 22.7\% & 99.4\%\\
\addlinespace
Indonesia & 277.6 & 4,139 & 1.5 & 13.0 & 4.7\% & 84.1\% & 27.5\% & 98.3\%\\
Afghanistan & 44.2 & 222 & 0.5 & 13.0 & 29.3\% & 39.1\% & 0.2\% & 60.8\%\\
Mozambique & 35.7 & 193 & 0.5 & 11.9 & 33.5\% & 30.9\% & 0.1\% & 93.6\%\\
Madagascar & 31.6 & 318 & 1.0 & 11.9 & 37.7\% & 33.7\% & 4.7\% & 81.3\%\\
Angola & 38.8 & 255 & 0.7 & 6.5 & 16.9\% & 56.9\% & 37.8\% & 88.5\%\\
\addlinespace
Brazil & 217.5 & 8,099 & 3.7 & 6.5 & 3.0\% & 89.5\% & 52.5\% & 97.6\%\\
Chad & 19.3 & 200 & 1.0 & 6.3 & 32.5\% & 20.1\% & 0.0\% & 82.6\%\\
Tanzania & 71.4 & 1,164 & 1.6 & 5.9 & 8.3\% & 52.1\% & 2.8\% & 97.7\%\\
North Korea & 26.0 & 54 & 0.2 & 5.8 & 22.5\% & 7.1\% & 1.4\% & 50.5\%\\
Niger & 29.2 & 152 & 0.5 & 5.8 & 19.9\% & 13.8\% & 0.0\% & 88.6\%\\
\addlinespace
Pakistan & 249.6 & 2,515 & 1.0 & 5.0 & 2.0\% & 79.9\% & 24.3\% & 96.4\%\\
Mali & 24.9 & 225 & 0.9 & 4.8 & 19.4\% & 21.8\% & 21.7\% & 94.1\%\\
Somalia & 15.5 & 56 & 0.4 & 4.3 & 27.9\% & 36.6\% & 0.0\% & 94.9\%\\
Mexico & 130.0 & 3,462 & 2.7 & 4.0 & 3.0\% & 86.7\% & 41.3\% & 97.9\%\\
Kenya & 57.1 & 1,106 & 1.9 & 3.8 & 6.7\% & 81.6\% & 59.8\% & 98.0\%\\
\bottomrule
\end{tabular}
\end{table}

Applying the Lancet Commission\textquotesingle s 80\%
threshold\textsuperscript{\citeproc{ref-meara2015}{6}} at 60 rather than
120 minutes, 148 of 162 countries reach it for any hospital, 101 for
emergency-tagged hospitals and 42 for tertiary-marker hospitals. The
geography changes with the tier (Figure \ref{fig:maps}b, c). Beyond 60
minutes of an emergency-tagged hospital, the uncovered population is no
longer rural and African: 22\% of India\textquotesingle s population,
44\% of Nigeria\textquotesingle s, 21\% of China\textquotesingle s and
16\% of Russia\textquotesingle s fall outside. Beyond 60 minutes of a
tertiary-marker hospital lives the majority of humanity, including 90\%
of India\textquotesingle s population, 74\% of
Bangladesh\textquotesingle s, 73\% of Indonesia\textquotesingle s, most
of sub-Saharan Africa, and about half of Brazil and Mexico.

\begin{figure}[H]

{\centering \includegraphics[width=1\linewidth]{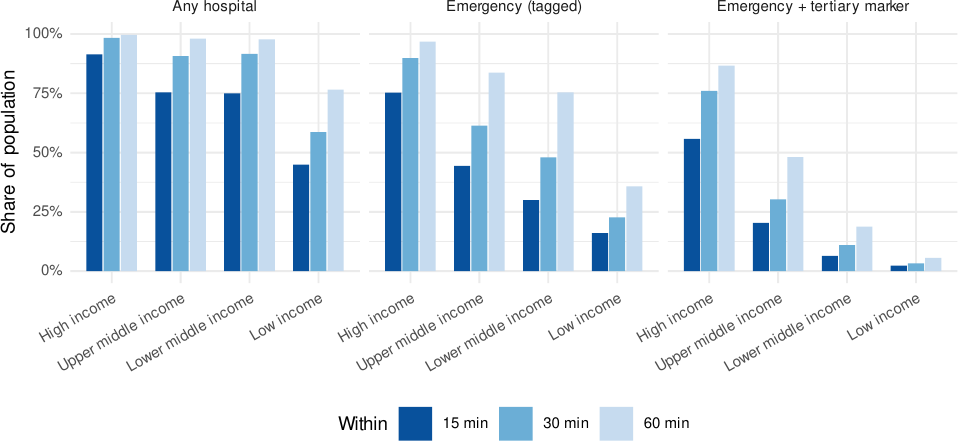} 

}

\caption{Share of population within 15, 30 and 60 minutes' drive of a hospital, by capability tier and World Bank income group (countries with at least one million inhabitants, population-weighted).}\label{fig:income-fig}
\end{figure}

By income group (Figure \ref{fig:income-fig}), any-hospital coverage at
60 minutes ranges from 99.7\% in high-income to 76.6\% in low-income
countries, and at 15 minutes from 91.4\% to 44.9\%. For the emergency
and tertiary tiers the gradient is steeper still: within 60 minutes of a
tertiary-marker hospital live 86.7\% of people in high-income, 48.1\% in
upper-middle, 18.8\% in lower-middle and 5.6\% in low-income countries.

\subsection{The capability gap and the documented-emergency
share}\label{sec:gap}

1.45 billion people (17.6\%) live within an hour of a hospital but not
of a hospital with a documented emergency department. India accounts for
311 million of this gap, China for 278 million, and Nigeria, Indonesia,
Pakistan, Russia and Japan for a further 187 million (Table
\ref{tab:gap}). In all of them only a small minority of mapped hospitals
carry the tag.

\begin{table}[!h]
\centering
\caption{\label{tab:gap}The fifteen countries with the largest capability gap: population within 60 minutes of any hospital but not of a hospital tagged with an emergency department.}
\centering
\fontsize{8}{10}\selectfont
\begin{tabular}[t]{lrrrrrrr}
\toprule
Country & \makecell[r]{Pop.\\(M)} & Hospitals & \makecell[r]{Tagged\\emergency} & \makecell[r]{Emergency\\share} & \makecell[r]{Within 60,\\any} & \makecell[r]{Within 60,\\emergency} & \makecell[r]{Capability\\gap (M)}\\
\midrule
India & 1,452.8 & 55,634 & 1,394 & 2.5\% & 99.9\% & 78.5\% & 310.7\\
China & 1,420.3 & 12,922 & 870 & 6.7\% & 98.4\% & 78.8\% & 278.0\\
Nigeria & 233.5 & 4,254 & 77 & 1.8\% & 92.6\% & 56.2\% & 85.0\\
Ethiopia & 132.2 & 405 & 18 & 4.4\% & 69.8\% & 23.3\% & 61.5\\
Pakistan & 249.6 & 2,515 & 167 & 6.6\% & 98.0\% & 79.9\% & 45.2\\
\addlinespace
Dem. Rep. Congo & 108.8 & 623 & 20 & 3.2\% & 61.4\% & 25.8\% & 38.7\\
Indonesia & 277.6 & 4,139 & 525 & 12.7\% & 95.3\% & 84.1\% & 31.1\\
Tanzania & 71.4 & 1,164 & 123 & 10.6\% & 91.7\% & 52.1\% & 28.2\\
Russia & 145.4 & 13,739 & 731 & 5.3\% & 99.0\% & 84.0\% & 21.8\\
Niger & 29.2 & 152 & 5 & 3.3\% & 80.1\% & 13.8\% & 19.4\\
\addlinespace
Myanmar & 54.3 & 2,532 & 81 & 3.2\% & 98.2\% & 63.2\% & 19.0\\
Uganda & 51.4 & 2,522 & 31 & 1.2\% & 99.6\% & 63.6\% & 18.5\\
North Korea & 26.0 & 54 & 0 & 0.0\% & 77.5\% & 7.1\% & 18.3\\
Sudan & 50.8 & 375 & 23 & 6.1\% & 61.9\% & 26.1\% & 18.2\\
Cameroon & 30.3 & 481 & 10 & 2.1\% & 95.4\% & 37.1\% & 17.6\\
\bottomrule
\end{tabular}
\end{table}

\begin{figure}[H]

{\centering \includegraphics[width=1\linewidth]{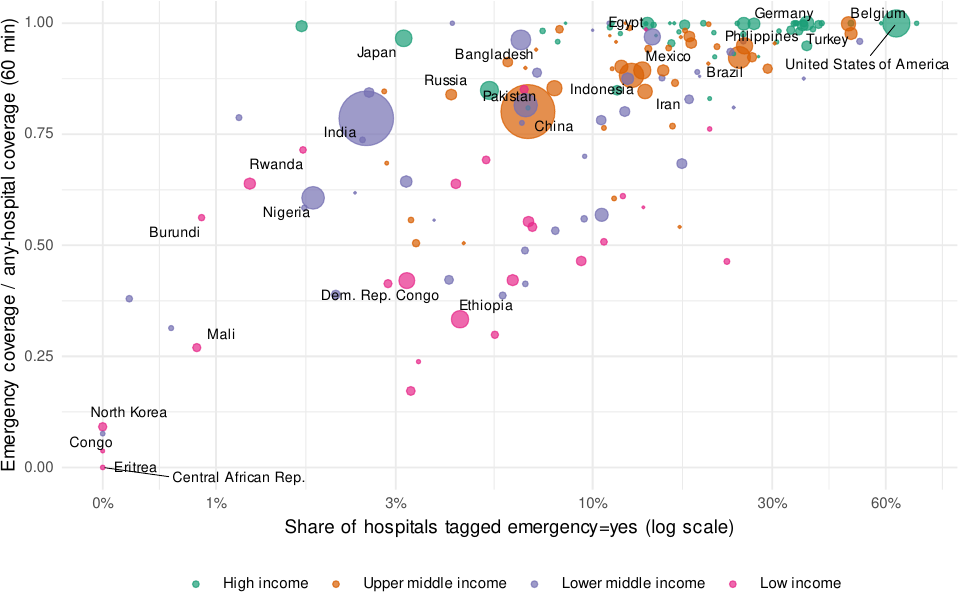} 

}

\caption{Ratio of 60-minute coverage by emergency-tagged hospitals to coverage by any hospital, against the share of a country's hospitals tagged emergency=yes. Countries with at least 20 mapped hospitals and one million inhabitants; point area proportional to population. Countries with no emergency-tagged hospital are plotted at 0.5\%.}\label{fig:gap-fig}
\end{figure}

Figure \ref{fig:gap-fig} plots, for countries with at least 20 mapped
hospitals, the ratio of emergency to any-hospital 60-minute coverage
against the documented-emergency share. Part of the relation is
mechanical: with no tagged hospital the ratio is zero, with all tagged
it is one. The informative part is the spread at low shares. Above a tag
share of about 20\% the ratio is close to one everywhere: the tagged
minority is spread so that it covers almost everyone the full set
covers. Below 5\% the ratio ranges from near zero to near one for
similar shares. At one end, the Central African Republic and Eritrea
have no tagged hospital and Congo and the Democratic
People\textquotesingle s Republic of Korea almost none, so their
near-zero emergency coverage says nothing about their hospitals. At the
other, India (2.5\% tagged, 78.5\% covered) shows that a small tagged
subset can be well distributed. Emergency-tier coverage should therefore
be read together with the documented-emergency share, and a share below
about 10\% treated as a signal that the tier reflects tagging rather
than capability.

\subsection{Benchmark against the friction-surface
product}\label{benchmark-against-the-friction-surface-product}

Re-weighted to the 2025 population grid, the friction-surface raster
places 96.7\% of the population within 60 minutes of a hospital or
clinic, 91.8\% within 30 and 75.0\% within 10 (98.9\% within two hours).
The 60-minute share exceeds the 91.1\% published in 2020; the most
likely reason is the population grid, since GHS-POP places people on
built-up cells that tend to lie near roads, and the choice of grid is
known to shift sub-national access statistics by tens of percentage
points\textsuperscript{\citeproc{ref-hierink2022}{38}}. Against this
benchmark the any-hospital estimate of 96.2\% is within one percentage
point, despite excluding clinics and routing only on drivable roads.
Cell by cell (Table \ref{tab:cross}), 7.67 billion people are within 60
minutes under both models, 208 million only under the friction surface,
167 million only under the network, and 98 million under neither.

\begin{table}[!h]
\centering
\caption{\label{tab:cross}Population (millions, GHS-POP 2025) cross-classified by the friction-surface travel time to any hospital or clinic (Weiss et al., 2020 raster) and by membership of the network-routed 60-minute polygons around any OpenStreetMap hospital.}
\centering
\fontsize{9}{11}\selectfont
\begin{tabular}[t]{lrr}
\toprule
  & Network: within 60 min of any hospital & Network: beyond 60 min\\
\midrule
Friction surface: 60 min or less & 7,665 & 208\\
Friction surface: more than 60 min & 167 & 98\\
Friction surface: no data & 47 & 7\\
\bottomrule
\end{tabular}
\end{table}

\begin{figure}[H]

{\centering \includegraphics[width=1\linewidth]{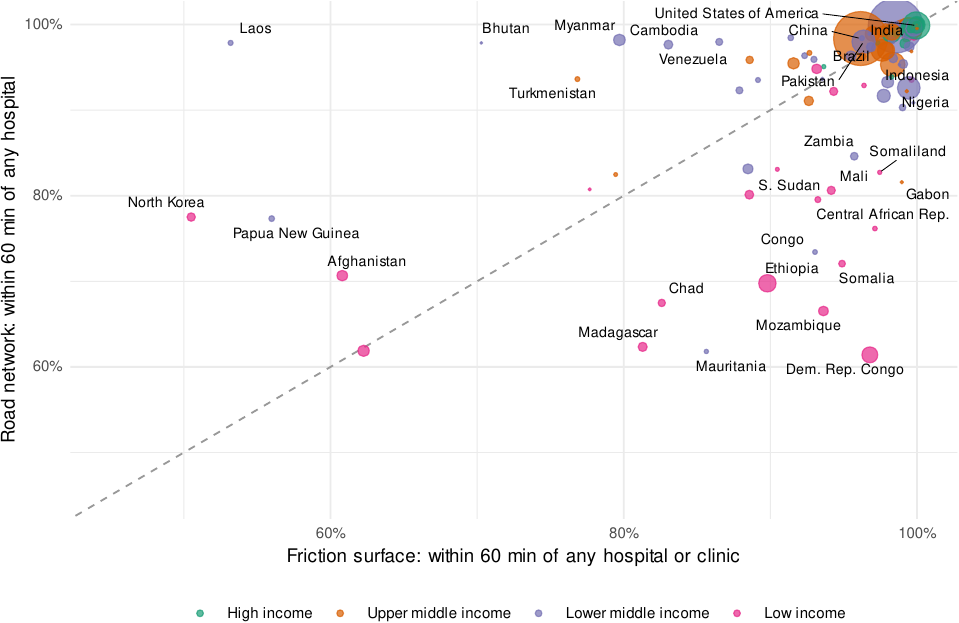} 

}

\caption{Country-level share of population within 60 minutes: network-routed drive time to any OpenStreetMap hospital (this study) against friction-surface travel time to any hospital or clinic (Weiss et al. 2020 raster), both weighted with GHS-POP 2025. Countries with at least one million inhabitants; point area proportional to population.}\label{fig:bench-fig}
\end{figure}

The comparison is between two products, not two methods in isolation:
facility definition (hospitals versus hospitals and clinics), access
model (network versus friction surface) and data vintage (2026 versus
2019) differ together, and the cross-tabulation cannot separate them.
Global agreement conceals systematic country-level disagreement (Figure
\ref{fig:bench-fig}). Across 160 countries the rank correlation between
the two estimates is 0.83 and the population-weighted mean difference
-0.5 percentage points, but the friction surface is more than five
points higher in 23 countries and more than five points lower in 11. The
first group is dominated by low-income African countries with sparse
road networks: friction surface versus network gives 96.8\% against
61.4\% in the Democratic Republic of the Congo, 89.8\% against 69.8\% in
Ethiopia and 93.6\% against 66.5\% in Mozambique. Three mechanisms
contribute: the friction surface counts clinics and health centres,
which in these countries outnumber hospitals many times over; it lets
travel cross any land cell, including tracks and footpaths excluded from
the car graph; and its facility data are from 2019 and partly from other
sources. The second group, where the network estimate is higher,
consists mainly of countries where OSM hospital mapping has grown since
2019 or where the friction surface assigns low speeds: Laos (53.2\%
against 97.9\%), the Democratic People\textquotesingle s Republic of
Korea, Bhutan, Papua New Guinea, Myanmar and Turkmenistan. Concave hulls
bridging unreached pockets between reached roads also inflate the
network estimate here, by up to a few percentage points of population
where roads are sparse (Section \ref{sec:sens}).

\subsection{Sensitivity to the isochrone construction}\label{sec:sens}

\begin{table}[!h]
\centering
\caption{\label{tab:sens}Sensitivity of population coverage to the isochrone construction. Columns 3 to 5: global coverage when every hospital's isochrone is the concave hull of reached road vertices at the stated ratio (1 would be the convex hull). Column 6: on the 2\% sample of hospitals (4,319 for any hospital, 516 emergency, 128 tertiary), the population inside the dissolved 1.5 km buffers around reached vertices divided by the population inside the dissolved 0.85 hulls of the same hospitals. Column 7: the main estimate multiplied by that ratio, i.e. the coverage implied if only people within 1.5 km of a reached vertex are counted.}
\centering
\fontsize{8}{10}\selectfont
\begin{tabular}[t]{llrrrrr}
\toprule
Tier & Band & Hull 0.85 (main) & Hull 0.5 & Hull 0.3 & Buffer / hull, sample & Buffer-equivalent\\
\midrule
 & 15 min & 75.2\% & 74.0\% & 72.5\% & 1.037 & 77.9\%\\

 & 30 min & 89.4\% & 88.7\% & 87.8\% & 0.976 & 87.2\%\\

\multirow[t]{-2}{*}[\normalbaselineskip]{\raggedright\arraybackslash Any hospital} & 60 min & 96.2\% & 95.9\% & 95.5\% & 0.946 & 91.0\%\\
\cmidrule{1-7}
 & 15 min & 42.0\% & 41.5\% & 40.9\% & 1.077 & 45.2\%\\

 & 30 min & 57.9\% & 57.2\% & 56.4\% & 0.993 & 57.5\%\\

\multirow[t]{-2}{*}[\normalbaselineskip]{\raggedright\arraybackslash Emergency (tagged)} & 60 min & 78.5\% & 77.8\% & 77.1\% & 0.967 & 76.0\%\\
\cmidrule{1-7}
 & 15 min & 19.8\% & 19.5\% & 19.2\% & 1.030 & 20.4\%\\

 & 30 min & 28.8\% & 28.5\% & 28.2\% & 0.957 & 27.5\%\\

\multirow[t]{-2}{*}[\normalbaselineskip]{\raggedright\arraybackslash Emergency + tertiary marker} & 60 min & 40.3\% & 39.8\% & 39.3\% & 0.941 & 37.9\%\\
\bottomrule
\end{tabular}
\end{table}

\begin{figure}[H]

{\centering \includegraphics[width=1\linewidth]{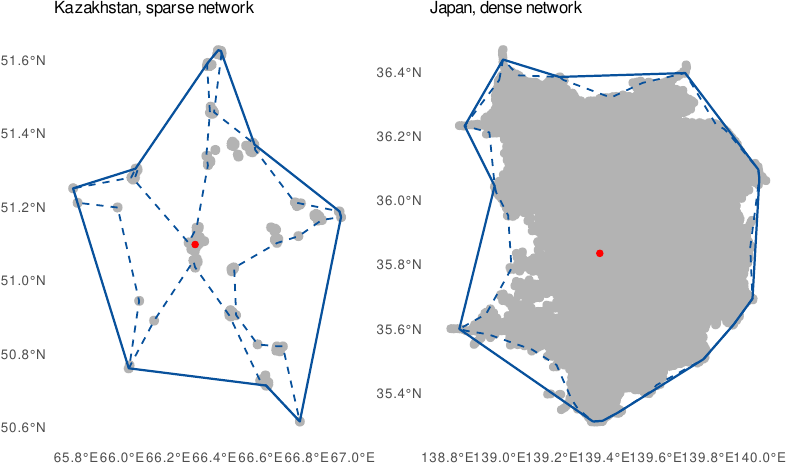} 

}

\caption{Isochrone construction for two sampled hospitals at 60 minutes. Grey fill: union of 1.5 km buffers around every road vertex reached within 60 minutes. Outlines: concave hulls of the same vertices at ratio 0.85 (solid, used in this study) and 0.3 (dashed). Left, a district hospital in Kazakhstan on a sparse network, where the hull covers 12 times the buffer area; right, a hospital in Japan on a dense network, where the ratio is 1.13.}\label{fig:sens-fig}
\end{figure}

The coverage shares are insensitive to the hull ratio (Table
\ref{tab:sens}). Making the hulls markedly more concave, from 0.85 to
0.3, lowers 60-minute coverage from 96.2\% to 95.5\% for any hospital,
from 78.5\% to 77.1\% for emergency hospitals and from 40.3\% to 39.3\%
for tertiary-marker hospitals; at 15 minutes the changes are somewhat
larger, up to 2.6 points. The hull ratio therefore matters far less for
people than for area: on the sample, the median 60-minute 0.85 hull is
1.41 times the area of the 1.5 km buffer (interquartile range 1.23 to
1.78) and 33\% of the hulls\textquotesingle{} summed area lies more than
1.5 km from any reached vertex, but only 5.9\% of the population inside
the hulls does, because people live along roads. The direction reverses
at 15 minutes: the hulls are smaller than the buffers (median area ratio
0.93) and credit 3.7\% fewer people, since a hull ends at the outermost
reached vertex while the buffer extends 1.5 km beyond it. Figure
\ref{fig:sens-fig} shows the two regimes: in a sparse network the hull
spans the wedges between a few long roads, in a dense one hull and
buffer nearly coincide.

Multiplying the main estimates by the sample\textquotesingle s
buffer-to-hull population ratios gives the coverage that would be
reported if only people within 1.5 km of a reached vertex were counted
(last column of Table \ref{tab:sens}): 91.0\% rather than 96.2\% within
60 minutes of any hospital, 76.0\% rather than 78.5\% for emergency
hospitals and 37.9\% rather than 40.3\% for tertiary-marker hospitals.
These are lower bounds under a strict reading of the road graph, and
conservative ones: in the full dissolve, neighbouring
hospitals\textquotesingle{} polygons cover part of the area that a
single hull adds beyond its buffer, so the sample ratios overstate the
correction. The truth lies between them and the hull estimates; the tier
ordering is unaffected and the gaps between tiers change by less than
three points. The tertiary-tier ratio rests on 128 sampled hospitals and
is indicative only. The inflation is largest where roads are sparse: on
the sample, the share of hull population beyond 1.5 km from a reached
vertex at 60 minutes is 1.9\% in high-income countries and 6.5\% in
low-income countries.

\section{Discussion}\label{discussion}

\subsection{What the tiers add}\label{what-the-tiers-add}

The reference figure of 91.1\% within an hour by motorised
transport\textsuperscript{\citeproc{ref-weiss2020}{12}} can be read as
evidence that geographic access to health care is largely solved outside
remote areas. For the lowest level of hospital care the present estimate
supports that reading: 96.2\% of people can, under free-flow conditions,
drive to some hospital within an hour and 89.4\% within half an hour. It
does not extend to time-critical care. Requiring a documented emergency
department removes 1.45 billion people from 60-minute coverage, and
requiring a tertiary marker leaves 40.3\% covered. Under-tagging
inflates the first drop, but the second rests on a helipad marker that
was systematically reviewed for every emergency-tagged hospital and that
gives the same coverage whether or not the emergency tag is required,
and neither can reverse the direction: the catchments relevant to
thrombectomy, major trauma or cardiac intervention belong to a few
thousand hospitals, not two hundred thousand, and most of humanity lives
more than an hour from them. National studies point the same way: direct
access to a thrombectomy centre within 30 minutes for 31\% of the United
States population\textsuperscript{\citeproc{ref-sarraj2020}{9}}, access
within an hour for 21\% of
India\textquotesingle s\textsuperscript{\citeproc{ref-asif2024}{10}},
and 29\% of sub-Saharan Africans more than two hours from a public
emergency hospital in 2015\textsuperscript{\citeproc{ref-ouma2018}{19}}.

\subsection{Network routing and friction
surfaces}\label{network-routing-and-friction-surfaces}

The second result is that the two products give near-identical global
totals and diverge systematically in the poorest countries. Because
facility set, access model and vintage are confounded, the country-level
divergence cannot be attributed to the access model alone; what follows
is the interpretation most consistent with its geography. Delamater and
colleagues\textsuperscript{\citeproc{ref-delamater2012}{14}} found in
Michigan that raster and network methods identify similar underserved
areas but different numbers of underserved people, with the raster
method more sensitive to speed settings and, there, the more
pessimistic. At planet scale the methods agree where roads are densely
mapped and disagree where they are not, and the friction surface is the
more optimistic. That follows from its construction: a population
cluster 20 km from the nearest mapped road is reachable across the
raster at walking or all-terrain speed, and unreachable on a car graph.
Petricola and
colleagues\textsuperscript{\citeproc{ref-petricola2022}{15}} likewise
traced a raster-network discrepancy of 300,000 people in flood-affected
Mozambique to incomplete road mapping, which affects the network
approach more. Which estimate is closer to the truth depends on whether
the missing roads exist. OSM road length was about 83\% complete
globally in 2017, with well-governed, well-connected countries best
mapped\textsuperscript{\citeproc{ref-barrington2017}{39}}, and
humanitarian mapping has added much since; but where roads are unmapped,
the friction surface\textquotesingle s assumption that every cell is
crossable at a landscape-dependent speed is also an assumption.
Replication in Lagos, where a friction surface gave a median of 5
minutes, a routing engine 11 minutes and actual driving 50
minutes\textsuperscript{\citeproc{ref-banke2021}{17}}, suggests that
both methods err on the optimistic side and that the free-flow times
used here are best-case values.

\subsection{The emergency tag as a data-quality
indicator}\label{the-emergency-tag-as-a-data-quality-indicator}

The third result concerns the data. The \texttt{emergency=yes} tag is
present on 11.7\% of hospitals globally and on 2.5\%, 3.1\% and 1.8\% of
hospitals in India, Japan and Nigeria. No plausible model of these
health systems produces that distribution; it reflects bulk imports of
hospital records without capability attributes, community conventions,
and the attention paid to locations over attributes. Any global analysis
of OSM emergency capability is therefore a joint analysis of capability
and tagging. The paper makes this explicit by reporting the
documented-emergency share beside every coverage figure and by showing
that above a share of roughly 20\% the tagged subset covers almost
everyone the full set covers. For users of OSM health data, the
emergency tier is a lower bound on access to emergency care, and
comparisons of it should be confined to countries with similar tag
completeness. For the OSM community and for the national facility master
lists now being built with WHO
support\textsuperscript{\citeproc{ref-whoghfd}{40},\citeproc{ref-macharia2025}{41}},
Table \ref{tab:gap} lists the countries where a campaign to record
emergency capability would resolve the status of hospitals serving
hundreds of millions of people.

\subsection{Limitations}\label{limitations}

The hospital inventory is OSM\textquotesingle s, whose completeness
varies by country, as shown for
buildings\textsuperscript{\citeproc{ref-herfort2023}{42},\citeproc{ref-zhou2022}{43}}
and for health facilities in
Africa\textsuperscript{\citeproc{ref-macharia2025}{41},\citeproc{ref-south2020}{44}},
and whose definition of a hospital varies with national usage and
contributor judgement: India\textquotesingle s 55,634 mapped hospitals
include many small private nursing homes, so its any-hospital coverage
is not comparable with that of countries where the tag is reserved for
larger institutions. Weiss and colleagues combined OSM with Google Maps
and published inventories; this study did not, because the tiers depend
on OSM attributes. Where OSM under-represents hospitals, any-hospital
coverage is underestimated; the divergence from the friction surface in
central Africa bounds the possible magnitude, though not all of it is
missing hospitals. The tertiary marker is a proxy: a university name
does not guarantee neurointerventional capability, a helipad does not
guarantee a trauma team, and some capable hospitals carry neither. The
helipad component is systematically complete for emergency-tagged
hospitals (Section 2.1), so within that set the tier is a fair census of
hospitals with helicopter access; the helipad tier shows that relaxing
the emergency requirement does not change its coverage, but helipads at
hospitals without the tag were not systematically reviewed, so that tier
may be incomplete, and the university-name component depends on naming
conventions and on the roots in the match list. Travel times assume
free-flow driving at default class speeds and an available vehicle; no
sensitivity analysis on speeds was performed, and the routed times have
not been validated against observed journeys, which in comparable
settings run longer than
modelled\textsuperscript{\citeproc{ref-banke2021}{17},\citeproc{ref-whitaker2022}{18}}.
Isochrones are near-convex hulls of reached vertices (ratio 0.85), which
enclose pockets not reachable within the budget and, in sparse networks,
bridge unreached terrain. Section \ref{sec:sens} bounds the effect:
relative to a 1.5 km buffer around reached vertices, the hulls overstate
60-minute population coverage by about 5\% for any hospital and 6\% for
tertiary-marker hospitals, and understate 15-minute coverage by a few
percent; the tier ordering is unaffected. Routing ignores borders, which
is realistic within the Schengen area and not elsewhere. Snapping
without a distance limit routed a few island hospitals from another
landmass, and 1,194 hospitals on disconnected vertices received no
isochrone. GHS-POP 2025 is a projection, and gridded populations differ
enough to move coverage statistics in sparsely populated
units\textsuperscript{\citeproc{ref-hierink2022}{38}}. Finally, polygon
membership is potential access: it says nothing about beds, staff,
supplies or cost, and patients often bypass the nearest
facility\textsuperscript{\citeproc{ref-bihin2022}{45}}.

\subsection{Future work}\label{future-work}

Buffered-edge isochrones or a rasterised travel-time surface would
remove the hull approximation and allow comparison of travel-time
distributions rather than band membership. Fusing OSM with open national
master facility
lists\textsuperscript{\citeproc{ref-maina2019}{21},\citeproc{ref-south2020}{44}}
would improve both the inventory and its attributes. Repeating the
analysis on successive planet files would separate change in access from
change in mapping, which OSM-based monitoring must control for.

\section{Conclusion}\label{conclusion}

Routing on the OpenStreetMap road network from 217,841 hospitals places
96.2\% of the world\textquotesingle s population within a 60-minute
free-flow drive of a hospital, within one percentage point of the
friction-surface estimate for any health facility when both use the same
2025 population grid. The agreement fails in the poorest countries,
where the friction surface is systematically more optimistic, and it
does not extend to time-critical care: 78.5\% of people live within an
hour of a hospital with a documented emergency department and 40.3\%
within an hour of one with a tertiary marker, a figure that is unchanged
when the tier is defined by a helipad alone without reference to the
emergency tag. Replacing the hull isochrones by a strict 1.5 km buffer
around reached road vertices lowers the 60-minute figures by 3\% to 6\%
of their value and leaves the tier ordering intact. The 1.45 billion
people between the first two tiers are concentrated in South and East
Asia and Nigeria, and part of that gap is in the data rather than the
health system. Access statistics that treat every facility alike
overstate access to the care the golden hour is about. Tiered, routed
estimates read together with an explicit indicator of attribute
completeness are a better basis for planning.

\section*{Data and code availability}\label{data-and-code-availability}
\addcontentsline{toc}{section}{Data and code availability}

The analysis code and the derived dataset are available at
\url{https://github.com/emailsson/global-hospital-travel-time}\textsuperscript{\citeproc{ref-repo}{46}},
with the larger polygon files as release assets from the same location,
all under the Open Database License as a derivative of OpenStreetMap.
The routing pipeline itself is not released; Section 2.2 specifies it in
sufficient detail to reproduce with open tools. Third-party inputs
(GHS-POP, Natural Earth, the World Bank income classification and the
Malaria Atlas Project raster) are available from their publishers.

\section*{Declarations}\label{declarations}
\addcontentsline{toc}{section}{Declarations}

\textbf{Funding.} This work received no specific funding and was carried
out in the author\textquotesingle s own time.

\textbf{Competing interests.} The author is employed by Safeture AB.
This work was carried out independently in the author\textquotesingle s
own time; the employer did not fund it and had no role in its design,
analysis or writing.

\textbf{Ethics.} The study uses only publicly available aggregate and
geographic data and involved no human participants; no ethics approval
was required.

\section*{Acknowledgements}\label{acknowledgements}
\addcontentsline{toc}{section}{Acknowledgements}

Map data copyright OpenStreetMap contributors, ODbL 1.0. The author
thanks the maintainers of osm2po, GDAL, GEOS, SciPy, GeoPandas, sf,
terra and R Markdown.

\section*{References}\label{references}
\addcontentsline{toc}{section}{References}

\protect\phantomsection\label{refs}

\end{document}